\documentclass{bmcart}

\usepackage{graphicx}
\usepackage[]{xcolor}
\usepackage{marvosym}
\usepackage{cite}
\usepackage{hyperref}
\usepackage{color} 
\usepackage{fancyhdr,graphicx} 

\expandafter\let\csname equation*\endcsname\relax

\expandafter\let\csname endequation*\endcsname\relax

\usepackage{amsmath}
\usepackage{mathrsfs}
\usepackage{float}
\usepackage[super]{nth}
\usepackage{amsfonts}
\usepackage{textcomp, gensymb}
\usepackage{multirow}
\usepackage{amssymb}
\usepackage{float,lscape}
\usepackage{url}
\usepackage{enumitem}
\usepackage{afterpage}
\usepackage{tabularx}
\usepackage{braket}
\usepackage{epstopdf}
\usepackage{lipsum}

\startlocaldefs
\endlocaldefs

\begin{document}

%%% Start of article front matter
\begin{frontmatter}

\begin{fmbox}
\dochead{Research}


\title{A blueprint for a simultaneous test of quantum mechanics and general relativity in a space-based quantum optics experiment}

%%%%%%%%%%%%%%%%%%%%%%%%%%%%%%%%%%%%%%%%%%%%%%
%%                                          %%
%% Enter the authors here                   %%
%%                                          %%
%% Specify information, if available,       %%
%% in the form:                             %%
%%   <key>={<id1>,<id2>}                    %%
%%   <key>=                                 %%
%% Comment or delete the keys which are     %%
%% not used. Repeat \author command as much %%
%% as required.                             %%
%%                                          %%
%%%%%%%%%%%%%%%%%%%%%%%%%%%%%%%%%%%%%%%%%%%%%%

\author[
   addressref={aff1},
   corref={aff1},
   email={sam.pallister@bristol.ac.uk}
]{\inits{SP}\fnm{Sam} \snm{Pallister}}
\author[
   addressref={aff3},
   email={simon.coop@icfo.es}
]{\inits{SC}\fnm{Simon} \snm{Coop}}
\author[
   addressref={aff5,aff20},
   email={valerio.formichella@polito.it}
]{\inits{VF}\fnm{Valerio} \snm{Formichella}}
\author[
   addressref={aff6},
   email={n.gampierakis@uea.ac.uk}
]{\inits{NG}\fnm{Nicolas} \snm{Gampierakis}}
\author[
   addressref={aff13},
   email={notaro.1448249@studenti.uniroma1.it}
]{\inits{VN}\fnm{Virginia} \snm{Notaro}}
\author[
   addressref={aff10},
   email={p.knott@sussex.ac.uk}
]{\inits{PK}\fnm{Paul} \snm{Knott}}
\author[
   addressref={aff14},                   % id's of addresses, e.g. {aff1,aff2}
%   corref={aff1},                       % id of corresponding address, if any
%   noteref={n1},                        % id's of article notes, if any
   email={}   % email address
]{\inits{RA}\fnm{Rui} \snm{Azevedo}}
\author[
   addressref={aff2},
   email={nikolaus.buchheim@mpq.mpg.de}
]{\inits{NB}\fnm{Nikolaus} \snm{Buchheim}}
\author[
   addressref={aff4},
   email={}
]{\inits{SdC}\fnm{Silvio} \snm{de Carvalho}}
\author[
   addressref={aff7, aff17},
   email={emilia.jarvela@aalto.fi}
]{\inits{EJ}\fnm{Emilia} \snm{J\"{a}rvel\"{a}}}
\author[
   addressref={aff8},
   email={}
]{\inits{ML}\fnm{Matthieu} \snm{Laporte}}
\author[
   addressref={aff9},
   email={jukka-pekka.kaikkonen@aalto.fi}
]{\inits{JPK}\fnm{Jukka-Pekka} \snm{Kaikkonen}}
\author[
   addressref={aff11},
   email={}
]{\inits{NK}\fnm{Neda} \snm{Meshksar}}
\author[
   addressref={aff12, aff17},
   email={timo.nikkanen@aalto.fi}
]{\inits{TN}\fnm{Timo} \snm{Nikkanen}}
\author[
   addressref={aff15},
   email={}
]{\inits{MY}\fnm{Madeleine} \snm{Yttergren}}
%\author[
%   addressref={aff16},
%   email={guenter.kargl@oeaw.ac.at}
%]{\inits{GK}\fnm{G\"{u}nter} \snm{Kargl}}
%\author[
%   addressref={aff18,aff21},
%   email={esteban.castro.ruiz@univie.ac.at}
%]{\inits{ECR}\fnm{Esteban} \snm{Castro-Ruiz}}
%\author[
%   addressref={aff18},
%   email={siddharth.koduru.joshi@oeaw.ac.at}
%]{\inits{SKJ}\fnm{Siddharth Koduru} \snm{Joshi}}
%\author[
%   addressref={aff19},
%   email={manuela.wenger@tugraz.at}
%]{\inits{MU}\fnm{Manuela} \snm{Wenger}}

%%%%%%%%%%%%%%%%%%%%%%%%%%%%%%%%%%%%%%%%%%%%%%
%%                                          %%
%% Enter the authors' addresses here        %%
%%                                          %%
%% Repeat \address commands as much as      %%
%% required.                                %%
%%                                          %%
%%%%%%%%%%%%%%%%%%%%%%%%%%%%%%%%%%%%%%%%%%%%%%

\address[id=aff1]{
\orgname{School of Mathematics, University of Bristol},
\city{Bristol}
\postcode{BS8 1TW},
\cny{UK}
}
\address[id=aff3]{
\orgname{ICFO-Institut de Ciencies Fotoniques, The Barcelona Institute of Science and Technology},
\street{08860 Castelldefels},
\postcode{}
\city{Barcelona},
\cny{Spain}
}
\address[id=aff5]{
\orgname{Politecnico di Torino},
\street{Corso Duca degli Abruzzi 24},
\postcode{10129}
\city{Torino},
\cny{Italy}
}
\address[id=aff6]{
\orgname{School of Natural Sciences, University of East Anglia},
\street{}
\city{Norwich}
\postcode{NR4 7TJ},
\cny{UK}
}
\address[id=aff13]{
\orgname{Sapienza University of Rome, Department of Mechanical and Aerospace Engineering},
\street{Via Eudossiana 18},
\postcode{00184}
\city{Rome},
\cny{Italy}
}
\address[id=aff10]{
\orgname{Department of Physics and Astronomy, University of Sussex},
\city{Brighton},
\postcode{BN1 9QH},
\cny{UK}
}
\address[id=aff14]{%                           % unique id
  \orgname{Faculdade De Ci\^{e}ncias da Universidade do Porto}, % university, etc
  \street{Rua do Campo Alegre 687},                     %
  %\postcode{4169-007}                                % post or zip code
  \city{Porto},                              % city
  \cny{Portugal}                                    % country
}
\address[id=aff2]{%
  \orgname{Max Planck Institute of Quantum Optics},
  \street{}
  \postcode{}
  \city{Garching},
  \cny{Germany}
}
\address[id=aff4]{
\orgname{University of Applied Sciences Wiener Neustadt, Aerospace Engineering Department},
\street{Johannes Gutenberg-Straße 3},
\postcode{2700}
\city{Wiener Neustadt},
\cny{Austria}
}
\address[id=aff7]{
\orgname{Aalto University Mets\"{a}hovi Radio Observatory},
\street{Mets\"{a}hovintie 114},
\city{Kylm\"{a}l\"{a}}
\postcode{02540},
\cny{Finland}
}
\address[id=aff8]{
\orgname{APC (AstroParticule et Cosmologie), Universit\'{e} Paris Diderot, CNRS/IN2P3, CEA/Irfu, Observatoire de Paris, Sorbonne Paris Cit\'{e}},
\postcode{F-75205}
\city{Paris}
\street{Cedex 13},
\cny{France}
}
\address[id=aff9]{
\orgname{Low Temperature Laboratory, Department of Applied Physics, Aalto University},
\street{PO Box 15100},
\postcode{FI-00076}
\city{Aalto},
\cny{Finland}
}
\address[id=aff11]{
\orgname{Albert Einstein Institute, Leibniz University Hanover},
\street{Callinstrasse 38},
\postcode{30167}
\city{Hanover},
\cny{Germany}
}
\address[id=aff12]{
\orgname{Radar and Space Technology Research Group, Finnish Meteorological Institute},
\street{PO Box 503},
\postcode{FI-00101}
\city{Helsinki},
\cny{Finland}
}
\address[id=aff15]{
\orgname{Chalmers University of Technology, Physics and Astronomy},
\street{Chalmersplatsen 4},
\postcode{412 96}
\city{G\"{o}teborg},
\cny{Sweden}
}
\address[id=aff20]{
\orgname{Istituto Nazionale di Ricerca Metrologica (INRiM)},
\street{Strada delle cacce 91}
\postcode{10135}
\city{Torino},
\cny{Italy}
}
\address[id=aff17]{
\orgname{Aalto University Department of Radio Science and Engineering, 13000},
\postcode{FI-00076}
\city{Aalto},
\cny{Finland}
}
%\address[id=aff16]{
%\orgname{Space Research Institute, Austrian Academy of Science},
%\street{Schmiedlstr. 6},
%\postcode{A-8042},
%\city{Graz},
%\cny{Austria}
%}
%\address[id=aff18]{
%\orgname{Vienna Center for Quantum Science and Technology (VCQ), Faculty of Physics, University of Vienna},
%\street{Boltzmanngasse 5},
%\postcode{A-1090}
%\city{Vienna},
%\cny{Austria}
%}
%\address[id=aff19]{
%\orgname{Institute of Communication Networks and Satellite Communications, Graz University of Technology},
%\street{Inffeldgasse 12},
%\postcode{8010}
%\city{Graz},
%\cny{Austria}
%}
%\address[id=aff21]{
%\orgname{Institute of Quantum Optics and Quantum Information (IQOQI), Austrian Academy of Sciences},
%\street{Boltzmanngasse 3},
%\postcode{A-1090},
%\city{Vienna},
%\cny{Austria}
%}

%%%%%%%%%%%%%%%%%%%%%%%%%%%%%%%%%%%%%%%%%%%%%%
%%                                          %%
%% Enter short notes here                   %%
%%                                          %%
%% Short notes will be after addresses      %%
%% on first page.                           %%
%%                                          %%
%%%%%%%%%%%%%%%%%%%%%%%%%%%%%%%%%%%%%%%%%%%%%%

\begin{artnotes}
%\note{Sample of title note}     % note to the article
%\note[id=n1]{Equal contributor} % note, connected to author
\end{artnotes}

\end{fmbox}% comment this for two column layout

%%%%%%%%%%%%%%%%%%%%%%%%%%%%%%%%%%%%%%%%%%%%%%
%%                                          %%
%% The Abstract begins here                 %%
%%                                          %%
%% Please refer to the Instructions for     %%
%% authors on http://www.biomedcentral.com  %%
%% and include the section headings         %%
%% accordingly for your article type.       %%
%%                                          %%
%%%%%%%%%%%%%%%%%%%%%%%%%%%%%%%%%%%%%%%%%%%%%%

\begin{abstractbox}

\begin{abstract}
In this paper we propose an experiment designed to observe a general-relativistic effect on single photon interference. The experiment consists of a folded Mach-Zehnder interferometer, with the arms distributed between a single Earth orbiter and a ground station. By compensating for other degrees of freedom and the motion of the orbiter, this setup aims to detect the influence of general relativistic time dilation on a spatially superposed single photon. The proposal details a payload to measure the required effect, along with an extensive feasibility analysis given current technological capabilities.
\end{abstract}

%%%%%%%%%%%%%%%%%%%%%%%%%%%%%%%%%%%%%%%%%%%%%%
%%                                          %%
%% The keywords begin here                  %%
%%                                          %%
%% Put each keyword in separate \kwd{}.     %%
%%                                          %%
%%%%%%%%%%%%%%%%%%%%%%%%%%%%%%%%%%%%%%%%%%%%%%


% MSC classifications codes, if any
%\begin{keyword}[class=AMS]
%\kwd[Primary ]{}
%\kwd{}
%\kwd[; secondary ]{}
%\end{keyword}

\end{abstractbox}
%
%\end{fmbox}% uncomment this for twcolumn layout

\end{frontmatter}

%%%%%%%%%%%%%%%%%%%%%%%%%%%%%%%%%%%%%%%%%%%%%%
%%                                          %%
%% The Main Body begins here                %%
%%                                          %%
%% Please refer to the instructions for     %%
%% authors on:                              %%
%% http://www.biomedcentral.com/info/authors%%
%% and include the section headings         %%
%% accordingly for your article type.       %%
%%                                          %%
%% See the Results and Discussion section   %%
%% for details on how to create sub-sections%%
%%                                          %%
%% use \cite{...} to cite references        %%
%%  \cite{koon} and                         %%
%%  \cite{oreg,khar,zvai,xjon,schn,pond}    %%
%%  \nocite{smith,marg,hunn,advi,koha,mouse}%%
%%                                          %%
%%%%%%%%%%%%%%%%%%%%%%%%%%%%%%%%%%%%%%%%%%%%%%

%%%%%%%%%%%%%%%%%%%%%%%%% start of article main body
% <put your article body there>

%%%%%%%%%%%%%%%%
%% Background %%
%%

%-----------------------------------------------------------------------------

\section{Introduction}\label{sec:introduction}

% Historical Background of GR and QM experiments
Since the beginning of the 20th century, quantum mechanics and general relativity have been the theoretical foundations of modern physics. However, while their predictive power has enabled us to understand the universe at both very small and very large scales, their validity has only been explored separately. Regarding general relativity, historic tests include measuring the perihelion of Mercury \cite{clemence1947relativity} and the Pound-Rebka experiment to quantify gravitational redshift \cite{pound1960apparent}. Even with the precision attainable with contemporary experiments, general relativity remains robust; for example, estimates of the Shapiro delay (the slowing of light as it passes by a massive body) have been acutely measured via radio links with the Cassini spacecraft \cite{Bertotti03}. Additionally, tests of the (weak) equivalence principle now concur with general relativity with accuracy $10^{-13}$ \cite{Will2006}, and will be bounded by $10^{-15}$ with the space-based MICROSCOPE experiment \cite{Dittus05}. This resilience to experimental analysis is also shared by quantum mechanics. The most stringent tests of quantum theory are precision tests of quantum electrodynamics (QED); these have now been verified to an accuracy of $10^{-12}$ using a one-electron quantum cyclotron \cite{odom2006new}.

% Conceptual differences and incompatibility between QM and GR
While quantum field theory, which is currently the best description we have of the quantum world, is fundamentally incompatible with general relativity, descriptions of quantum field theories on static but curved background spacetimes are internally consistent and produce novel predictions \cite{Birrell1984} (such as Hawking radiation \cite{hawking1975particle}). However, any full description of gravity itself as a quantum field theory is bound to fail due to a lack of renormalisability. Additionally, quantum field theory and general relativity have contrasting descriptions of physical parameters, the most notable of which is the nature of \emph{time}; this ``problem of time'' is an unresolved hurdle in producing a canonically quantised theory of gravity \cite{Isham92}.

% Motivation of proposal by Zych et al.
Without exception, all of the experiments mentioned above are tests of either general relativity or quantum mechanics, but not the interplay of both. As such, their outcomes are consistent with classical dynamics evolving on a curved background (in the former case), or of a quantum field theory evolving in flat space (in the latter case). A controlled experiment exploring quantum dynamics on a curved background is, to the authors' knowledge, yet to be performed. To this end, a theoretical proposal by Zych et al. \cite{Zych2012} outlined the premise of a controllable experiment on a system that is both highly quantum, and occurring on a non-trivial background spacetime. Such an experiment would provide vital evidence to aid the successful unification of quantum mechanics and general relativity.

The crux of the proposal in \cite{Zych2012} is to perform interferometry of single quanta, but to orient the interferometer in such a way that quantum states in each arm traverse a different gravitational equipotential. Historically, such an idea is not new; the analysis of gravitational effects on matter interferometry was first explored by Collela, Overhauser and Werner (``COW'') in 1975 \cite{Colella1975}. While the COW experiment did observe a gravitational influence on the output of the interferometer, their data is adequately described by analysing a Mach-Zehnder interferometer placed in a \emph{Newtonian} gravitational field, rather than a truly relativistic spacetime. The key physical extension of the COW experiment present in \cite{Zych2012} is to construct an interferometer large enough that general relativistic effects become apparent in the output of the interferometer, and to let the position of the photon in the interferometer serve as the local clock. Similar proposals exploring gravitational effects on quantum optical systems consider interferometry with a precessing polarisation serving as the clock \cite{Brodutch2015}, or explore the degradation of entanglement due to gravity \cite{Bruschi2014}.

\begin{figure}[t!]
\centering
    \includegraphics[width=0.6\linewidth, trim = 0cm 0cm 0cm 0cm]{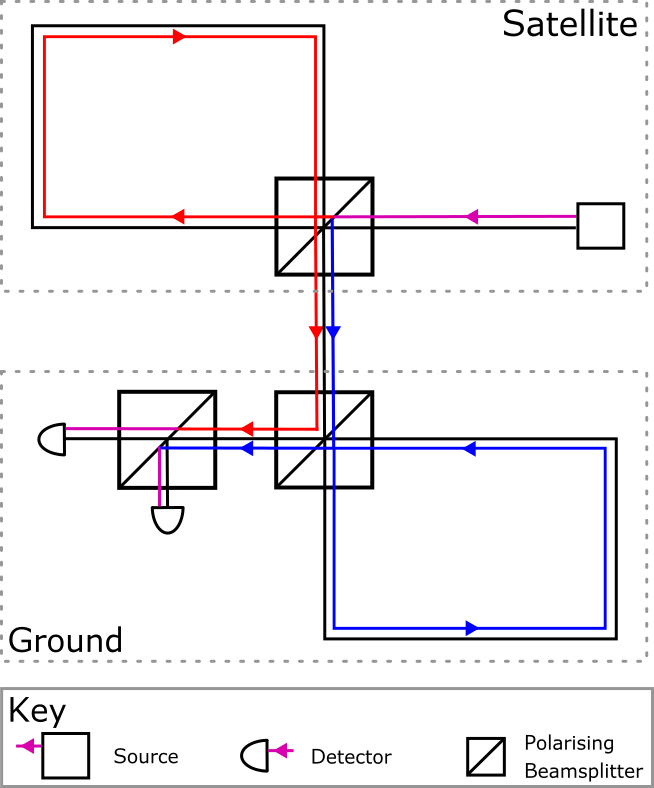}
    \caption{\csentence{A simplified schematic of the proposed experiment.} Colour here does not represent frequency; red and blue denote orthogonal polarisation states. The red and blue paths are only distinguished by their travel times around the loops on the satellite and on the ground, respectively.}
    \label{fig:setup_simpler}
\end{figure}

% Objective of the study
The objective of this paper is to present a full design of a space-based quantum optics experiment capable of carrying out the experiment proposed by Zych et al. \cite{Zych2012}, along with a discussion of the practical feasibility of such an experiment. The design consists of a ``folded'' Mach-Zehnder interferometer distributed between an orbiting satellite (containing a single photon source, a spool of single-mode optical fibre, and a transmitting telescope) and a ground-station (equipped with a receiving telescope, an identical spool of fibre and a bank of single photon detectors). A stripped-down schematic of the proposed design is shown in Figure~\ref{fig:setup_simpler}. Moreover, we contend that overcoming the engineering requirements for this proposal would be of significant benefit to those interested in developing quantum communications networks on a global scale.

The paper breakdown is as follows: in Section~\ref{sec:scientific_background}, we extend the analysis present in \cite{Zych2012} to derive the predicted interferometric signal given that the satellite is in motion. In Section~\ref{sec:implementation}, we give an extensive feasibility analysis of the experiment. This includes a proposed setup and operational parameters for key optical components in Section~\ref{sec:setup}, followed by analysis of interferometric stability, systematic error, loss and noise in Sections~\ref{sec:stability},~\ref{sec:dispersion},~\ref{sec:loss} and~\ref{sec:noise}, respectively. All of these elements are combined with a statistical analysis in Section~\ref{sec:hypothesis_testing}, in order to derive the operational duration necessary to test the hypotheses in \cite{Zych2012} to a prespecified confidence. Finally, we outline orbit and mission requirements in Section~\ref{sec:mission_design} and highlight operational risks to be mitigated in Section~\ref{sec:risk}.

%----------------------------------------------------------------------------------

\section{Scientific background}\label{sec:scientific_background}

While the experiment proposed by Zych et al. in \cite{Zych2012} calculates the general relativistic effect on the fringe visibility of the interferometer output signal, we must also introduce special relativistic corrections due to the orbital velocity of the satellite (Section~\ref{sec:time_dilation}). This full expression for the time dilation is then used to derive an expected output for the Mach-Zehnder interferometer in a regime where relativistic corrections are important (Section~\ref{sec:interferometry}).

\subsection{Relativistic time dilation}\label{sec:time_dilation}
General relativity predicts that a clock in a lower gravitational potential ticks slower than a clock in a higher one. This means that two clocks initially synchronised and then moved to points of differing gravitational potential accumulate an increasing time delay. This gravitational time dilation is observed daily in global navigation satellite systems, where one also has to account for special relativistic time dilation due to the motion of the positioning satellites. The total time delay can be derived directly in a general relativistic framework starting from the spacetime metric, using the weak field approximation for the Earth's gravitational potential and assuming the speed of a satellite to be much smaller than the speed of light.

It is important to note that the time dilation is independent of the physical realisation of the clock. For example, we could consider an optical fibre loop and use the position of a photon travelling in the fibre as a clock. If two of these fibre-clocks were placed at different heights in the Earth's gravitational field, the one at the lower position would tick slower, meaning that the photon would take more time to travel the entire length of the fibre. Such a time delay is known as the Shapiro delay.

\begin{figure}
\centering
 \includegraphics[width=0.8\linewidth, trim = 2cm 0cm 2cm 0cm]{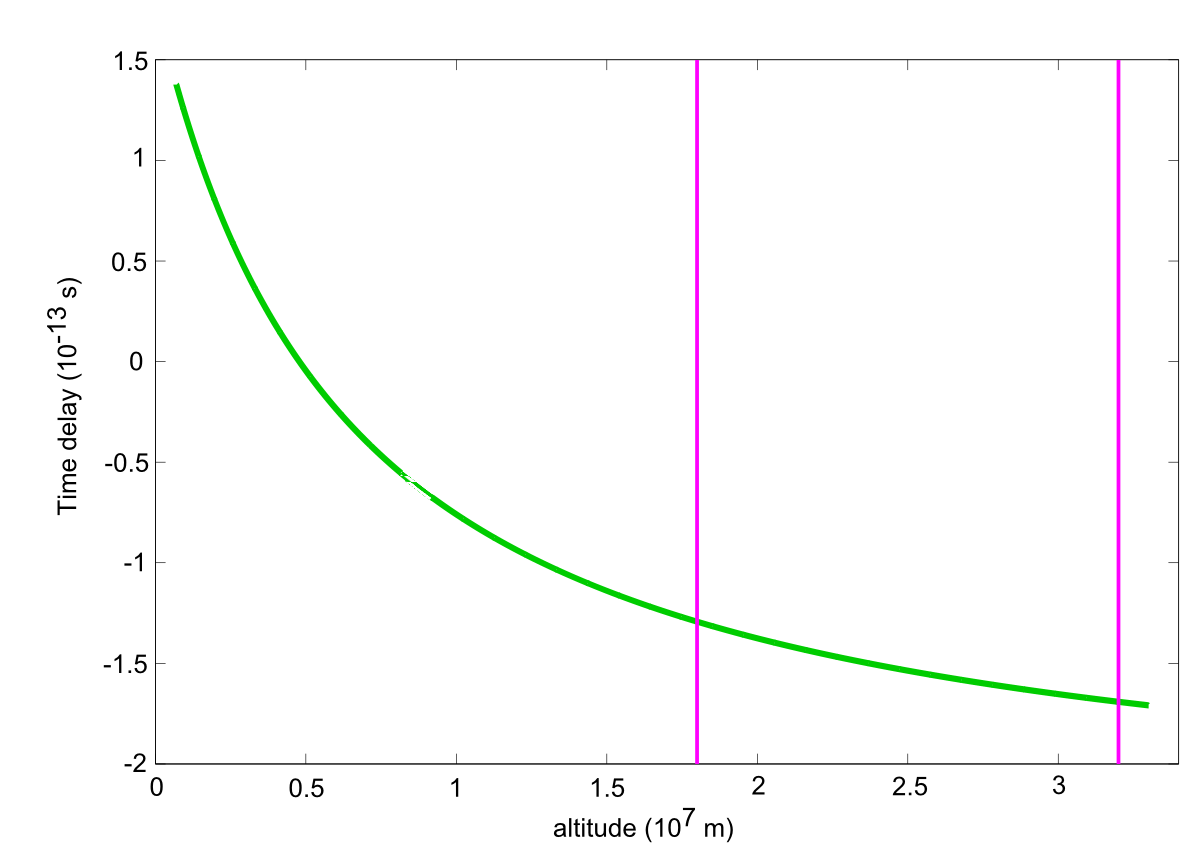}
 \caption{\csentence{The total time delay between the fibre clocks, as a function of height, for a fixed fibre length and fibre refractive index.} Magenta lines mark lowest and highest altitudes at which measurements could be performed (see feasibility analysis). In this region the time delay is dominated by the Shapiro delay. The choice of fibre length and refractive index is discussed in Section~\ref{sec:implementation}. The altitudes and orbital parameters are discussed in Section~\ref{sec:mission_design}.}
 \label{fig:timedelay}
\end{figure}

Considering the simplified experimental setup described in Figure~\ref{fig:setup_simpler}, we deduce that the time spent by a photon in the satellite loop differs from the time spent in the ground loop, as measured by an observer on the ground. The accumulated time delay, i.e. the difference in photon arrival times, can be obtained by generalising the calculation described in \cite{Zych2012} (see Appendix~\ref{appendix:deltatau}). For a satellite on an elliptical orbit around the Earth, the time delay is given by
\begin{equation}
\Delta \tau = \frac{n l}{c^3} \left[ -W_0 + GM \left( \frac{2}{R_{\oplus}+h} - \frac{1}{2a} \right)  \right]-\frac{n dl}{c} \label{eq:tdelay}
\end{equation}
where $l$ is the proper length of the satellite fibre, $l+dl$ is the proper length of the ground fibre, $n$ is the fibre's refractive index, $G$ is the gravitational constant, $M$ is the mass of the Earth (5.97219 $\times 10^{24}$\,kg), $R_{\oplus}$ is the radius of the Earth (6378100\,m), the geoidal potential $W_0=L_Gc^2$ where $L_G=6.969290134\times 10^{-10}$ by definition, $c$ is the speed of light in vacuum, $h$ is the altitude of the satellite with respect to the Earth's surface and $a$ is the semi-major axis of the orbit. In order to compute the gravitational potential experienced by the satellite, we assumed the Earth is a perfect sphere. We can safely neglect the corrections due to the irregular shape of the Earth, both on the gravitational potential and on the satellite's orbit, since they are orders of magnitude smaller than the first order approximation \cite{Kouba}. A plot of this expression, in the case $dl=0$, is shown in Figure~\ref{fig:timedelay}. Note that the general relativistic and special relativistic time dilations act in opposition, such that at low satellite altitudes the time delay is positive and dominated by the special relativistic dilation, whereas at high altitudes it is negative and dominated by the general relativistic dilation. Therefore care must be taken to launch the satellite into a sufficiently high orbit for the general relativistic time dilation to be the dominant effect. Figure~\ref{fig:timedelay} demonstrates that the Shapiro delay for a satellite at the orbital heights considered in this proposal is of the order $100$\,fs.

\subsection{Interferometry in the presence of gravity}\label{sec:interferometry}

\begin{figure}
\centering
\includegraphics[width=0.9\linewidth,trim = 1.9cm 0.5cm 1.9cm 0.9cm]{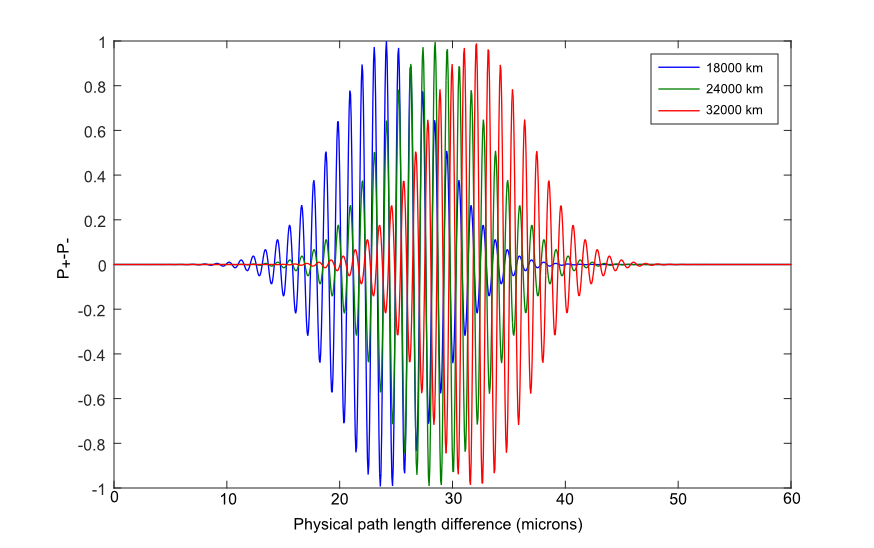}
 \caption{\csentence{The expected interferometer signal vs. physical path length difference of the two interferometer arms, with the parameters as described in the text, at three different heights of the satellite.} The signal was calculated by numerically evaluating Eq. (\ref{eq:MZ}) with a frequency-dependent refractive index as described in \cite{Malitson65}.}
 \label{fig:fringe_contrast}
\end{figure}

Given this modification to the expected time dilation, we provide a brief exposition on the theory presented in \cite{Zych2012}, in order to motivate the design of our experiment. The reader is referred to \cite{Zych2012}, Sections 2 and 3 for a more thorough treatment.

The output of a Mach-Zehnder interferometer is sensitive only to a difference in the relative optical path length of each arm \cite{Loudon2003}. If the interferometer were to have two ideally-controlled arms of equal proper length (i.e. $dl=0$), the relative optical path would be dependent only on the Shapiro delay $\Delta \tau$. However, by introducing a controlled difference in the relative length of the two interferometer arms by actively controlling the length of the fibre spools, one could cancel out $\Delta \tau$ and recover maximal contrast at the output of the interferometer for any height of the satellite. Measuring the path-length difference that gives maximum contrast at different gravitational potentials would constitute a measurement of the Shapiro delay between the parts of the photon state traversing the upper and lower arms of the interferometer.

It can be shown (see Appendix~\ref{appendix:signal}) that the probability of the photon being detected at the $\pm$ outputs of the Mach-Zehnder interferometer (with a single photon input) are given by:
\begin{equation}
P_\pm = \frac{1}{2} \left( 1 \pm \int d\nu |f(\nu)|^2 \textrm{cos}(\nu \Delta \tau ) \right),
\label{eq:MZ}
\end{equation}
where $\nu$ is angular frequency in rads$^{-1}$ defined in the reference frame of the observer at distance $r$ from the Earth, $\Delta \tau$ is the time delay as given in Eq. (\ref{eq:tdelay}), $P_{\pm}$ is the probability of the photon being detected at the $\pm$ output, and $f(\nu)$ is the spectrum of the light source, normalised such that $\int^\infty_{-\infty} |f(\nu)|^2 d\nu = 1$.

In Figure \ref{fig:fringe_contrast} we numerically evaluated Eq. \ref{eq:MZ} with a range of different fibre lengths for three different satellite heights. We assumed a Gaussian spectral density for the light with $f(\nu) = \left( \frac{\sigma}{\pi} \right)^{1/4} \textrm{exp}(-\frac{\sigma}{2}(\nu - \nu_0)^2)$, with parameters taken from the setup in Section~\ref{sec:implementation}: a central frequency $\nu_0 = 2 \pi \times c /$1550\,nm and width $\sigma = (100\,\textrm{ fs} / 2 \pi)^2$. We take the satellite fibre to be 60\,km long, while the length of the ground station fibre differs by the amount given on the x-axis of the figure. The frequency-dependent speed of light in the fibres was calculated from the refractive index for fused silica given in \cite{Malitson65}.

In our experiment the satellite is sent on an elliptic orbit and we record the physical path-length difference required to recover full visibility, for a range of satellite heights. From this, the corresponding Shapiro delay as a function of satellite height can be inferred.

%----------------------------------------------------------------------------------

\section{Implementation}\label{sec:implementation}

\begin{center}
\begin{figure*}
\hspace*{2.2cm}
 \includegraphics[width=0.75\textwidth, trim = 3.7cm 0cm 0cm 0cm]{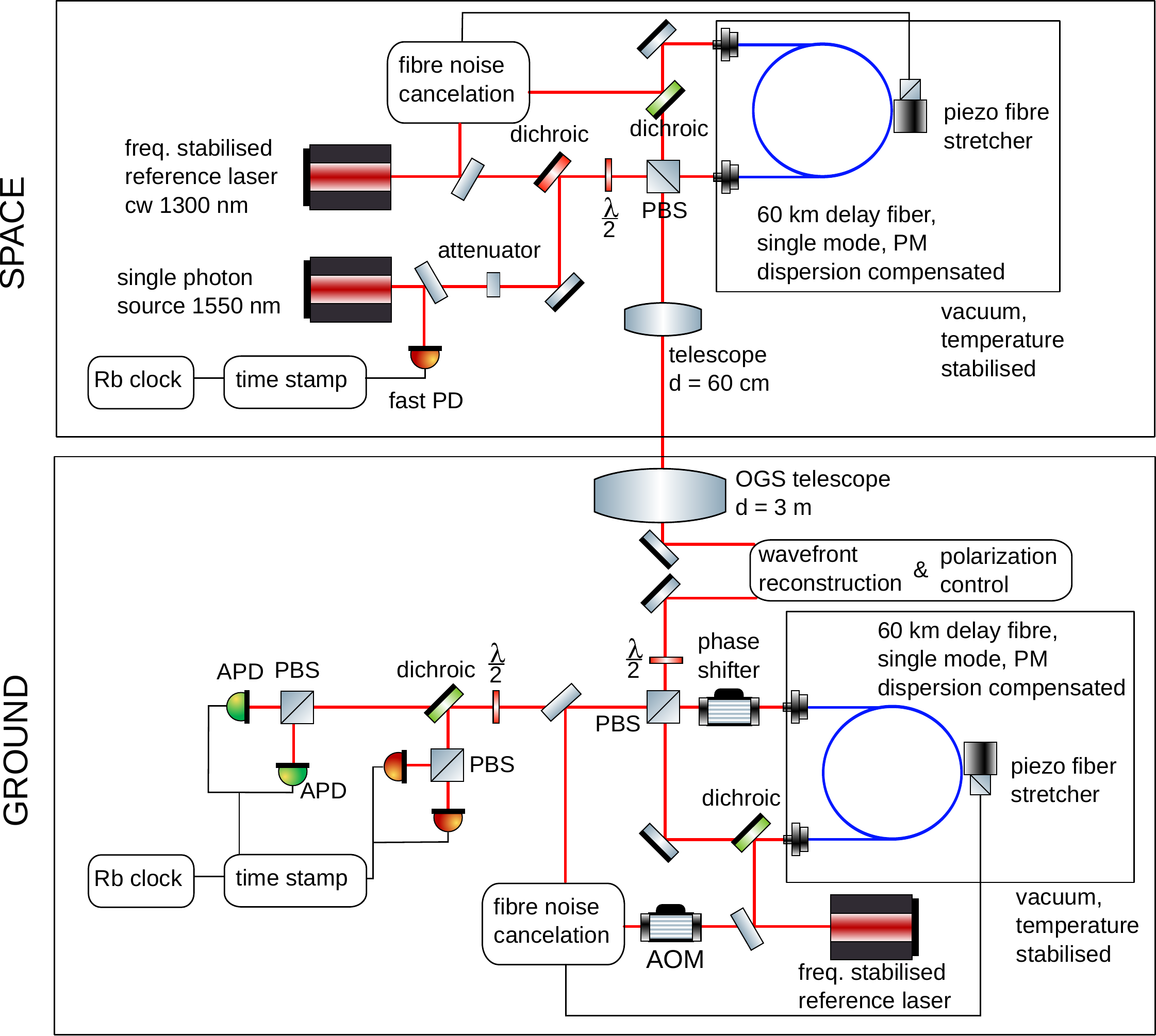}
 \caption{\csentence{Full schematic of the payload optical system.} The setup is designed to minimise the effects of other degrees of freedom besides the Shapiro delay that may introduce distinguishability at the output of the interferometer.} %update to the new setup figure. ground.png and space.png which are located in dropbox/Team Blue/AAA Post-A presentation parts.
 \label{fig:setup}
\end{figure*}
\end{center}

In order to present a feasible implementation, it must be demonstrated that enough data can be collected within the lifetime of the satellite to confirm, or refute, the hypothesised signal to a pre-specified confidence. In this particular proposal, there are four key issues that must be addressed before feasibility can be demonstrated: stability of the interferometer, loss from the source to the detectors, noise from extraneous and atmospheric photons, and mitigation of any other degrees of freedom besides the Shapiro delay that might introduce distinguishability (in particular, photon dispersion). We first outline a feasible setup in Section~\ref{sec:setup}, before presenting an analysis of these key issues.

% Setup ----------------------------------------------------------------------------------
\subsection{Optical setup and components}\label{sec:setup}

A detailed illustration of the interferometer setup, with all the key optical components, can be seen in Figure~\ref{fig:setup}.

The choice of operational parameters is ultimately a compromise between what is feasible to implement and what is necessary to recover the desired data. As such, there is no single optimal set of components; relaxing the constraints on their performance may be acceptable if they become more durable and the satellite can operate for longer and take more data, for example (we aim to mitigate large pulse dispersion, noise and random fluctuations in signal by integrating counts over long time periods; hence, durability of components is critical). Therefore, the following choices of components and parameters are not intended to be a fixed specification for the payload, but to demonstrate one setup that could feasibly produce meaningful results.

We highlight the following optical elements, whose operational parameters are crucial to the success of the experiment.

\emph{Single photon source:} An off-the-shelf, mode-locked, pulsed laser with operational wavelength of 1550\,nm and 1\,GHz repetition rate is attenuated to a mean photon number of  0.1 photons per pulse. While an attenuated laser is a straightforward option, true, heralded single photon sources have been space-qualified and operated on the QUESS satellite \cite{Gibney2016}. However, we are not aware of a true, heralded source of single photons that meets the requirements of repetition rate and pulse width presented here. The reason for this particular level of attenuation is the tradeoff between lowering multiphoton emission and maintaining a high number of total counts - this attenuation reduces the probability of multiphoton emission to 5\% and results in a single photon rate of 100\,MHz, which matches the resolution of the single photon detectors. The chosen wavelength guarantees high atmospheric transmission and the utilisation of well-developed telecommunication technology. The functional dependence of the fringe contrast on the Shapiro delay $\Delta \tau$ and the pulse width $\sqrt{\sigma}$ (see \cite{Zych2012}, Eq. 13) demands the utilisation of ultra-short ($<$1\,ps) single photon pulses to measure a noticeable effect (we note single photon sources with a width as short as 65\,fs have been demonstrated in \cite{Mosley2008}). We take $\sqrt{\sigma} = 150\,\text{fs}$.

\emph{Classical reference laser:} A 1300\,nm multi-purpose laser, operating in continuous wave mode, is led alongside the single photons through the interferometer. Its purpose is to provide reference data to estimate phase fluctuations and systematic errors. For precise corrections, an operational wavelength as close as possible to the single photon source is required, but overlap with the bandwidth of the single photon pulses has to be avoided.  The chosen wavelength is a reasonable compromise. Frequency stability of both lasers is paramount; see Section~\ref{sec:stability}. Additionally, a fraction of the incident laser power is separated at the ground station and used for wavefront reconstruction and compensation of polarisation changes due to satellite movement.

\emph{Fibres:} We expect the delay fibres must be stabilised to relative length changes of $10^{-10}$ for observing the predicted interference effects, which includes a thermal stabilisation of $\pm 10^{-5}$\,K. Moreover, the fibre length must change dynamically to recover full contrast of the interference fringes. Active length corrections are carried out by a piezo fibre-stretcher. Fibres of length $l = 60\,\text{km}$ and refractive index $n = 1.5$ (glass) are assumed.

\emph{Transmission telescope:} The onboard emitting telescope is used to focus the beam from the two sources. The aperture needs to be large enough to emit a strong signal but is limited by size and weight; a reasonable choice is a 60\,cm aperture, 1\,m long, 6\,$\mu$rad field of view, Cassegrain reflector telescope. Material choices, such as Beryllium mirrors, would also minimise weight.

\emph{Single photon detectors:} Since the goal of the detectors is to receive single photons that have travelled astronomical distances, it is extremely beneficial to minimise dark counts and maximise efficiency. We consider single-nanowire single photon detectors (SNSPDs), to benefit from the superior dark count over conventional avalanche photodiodes. While these detectors aren't as commonplace, the technology has already been established in a similar setting in the LLCD mission (NASA) \cite{Biswas2014}.

\emph{Timing:} Emission of the single photon pulses are tagged with timestamps by a Rb-clock (chosen primarily for its small size, weight and commercial availability). Synchronisation with an identical clock on the ground will be used to exclude background noise at the data post-processing stage and to track the path of the satellite. Precision timing is also necessary to modulate the action of time-gate filters in front of the detectors.

  %With these experimental parameters, we calculate time delays between -130 fs (lowest altitude) and -170 fs (highest altitude). The predicted time delay as a function of the satellite altitude was shown above, in Figure~\ref{fig:timedelay}. The change in the fringe contrast along the orbit is shown in Figure~\ref{fig:visibility} ({\color{red} to be substituted}).

% Stability ------------------------------------------------------------------------------
\subsection{Random fluctuation and stability}\label{sec:stability}

As the effect we wish to measure is a minute change in optical path length in the interferometer due to the Shapiro delay, exposing the interferometer to other sources of instability can swamp the desired signal. We find that optical path and phase changes in the atmosphere due to temperature fluctuations are negligible compared to fluctuations in the length of fibre; given the thermal properties of fused silica we calculate necessary temperature stabilisation of the fibres of less than $10^{-5}$\,K to ensure a relative path length stability of $10^{-10}$. Passive insulation and active heat distribution can mitigate a large fraction of the thermal instability, but both the satellite and the ground station include a feedback loop of a frequency stabilised reference laser in combination with a piezo fibre stretcher, to allow for fibre noise reduction. Also, some thermal fluctuation can be erased in post-processing by referring to data from the reference laser.

In addition, continued operation of the experiment requires protection of the reference laser from frequency fluctuations and drift. Given the magnitude of the path length change, we estimate a necessary relative frequency stability of the reference laser less than $10^{-11}$. Long term accuracy of the reference laser within required precision is most feasible via frequency comb stabilisation \cite{Ferreiro2011}. This method will increase further complications and costs, and could be circumvented by improvements in the area of stabilisation by using atomic absorption lines, which today are close to reaching comparable relative accuracies \cite{Jiang2005}.

% Dispersion -----------------------------------------------------------------------------
\subsection{Systematic transmission errors and dispersion}\label{sec:dispersion}

We highlight three systematic errors present in the experiment, mitigation of which are critical: dispersion, both in the fibre and the atmosphere; Doppler shift due to the velocity of the satellite; and changes in optical path length due to the ellipticity of the orbit.

Dispersion is prevalent both in the optical fibre and in the atmosphere. We estimate a requirement for fibre dispersion of $<$\,5\,fs/km/nm, ensuring a broadening of the pulse width in the fibres of $<$\,0.5\% per km; this is a stringent enough requirement to fix dispersion as the primary hurdle facing this experiment. The current state-of-the-art for dispersion-limited fibre is a factor of ten worse than this: 50\,fs/km/nm \cite{Shen2003}. Current technologies are capable of dispersion compensation in optical fibres of 0.5\,ps/km/nm, which has to be improved by about a factor of ten to make the scientific requirement feasible. However, the utilisation of telecommunication fibre wavelengths ensure ongoing research in that scientific area. For example dispersion-free transmission of 610\,fs laser pulses over a 160\,km fibre has been demonstrated, but dispersion-free transmission techniques typically suffer from large losses \cite{Tologlou2011} \cite{Pelusi2000}. On the other hand active compensation of \emph{atmospheric} chromatic dispersion of a 250\,fs pulse over a 200\,km propagation distance, to a uncertainty of $\pm$10\,fs, has been demonstrated \cite{Lee11}. This indicates that control of atmospheric dispersion of comparably short pulses is well within reach, especially considering the pulses in this experiment are travelling vertically, on a much shorter trajectory through the atmosphere.

Doppler shift, conversely, presents much less of a problem. This is straightforward to calculate based on the velocity of the satellite relative to the Earth. For the worst possible case (at apogee), the Doppler redshift $\nu$ is calculated using the formula:

\begin{equation}
\Delta\nu_{Doppler} =f_p\times \frac{c}{c-v_0}
\end{equation}

where $c$ is the speed of light, $f_p$ is the frequency of the single photon source and $v_0$ is the maximum velocity of the satellite at the lowest sampling point. This gives a shift in photon frequency of 2.48\,GHz. Considering the bandwidth size and that the photon frequency exceeds 193\,THz, this is considered negligible.

\begin{figure}
  \hspace*{0.5cm}
  \includegraphics[width=0.7\linewidth, trim=1.2cm 0.5cm 0cm 0cm]{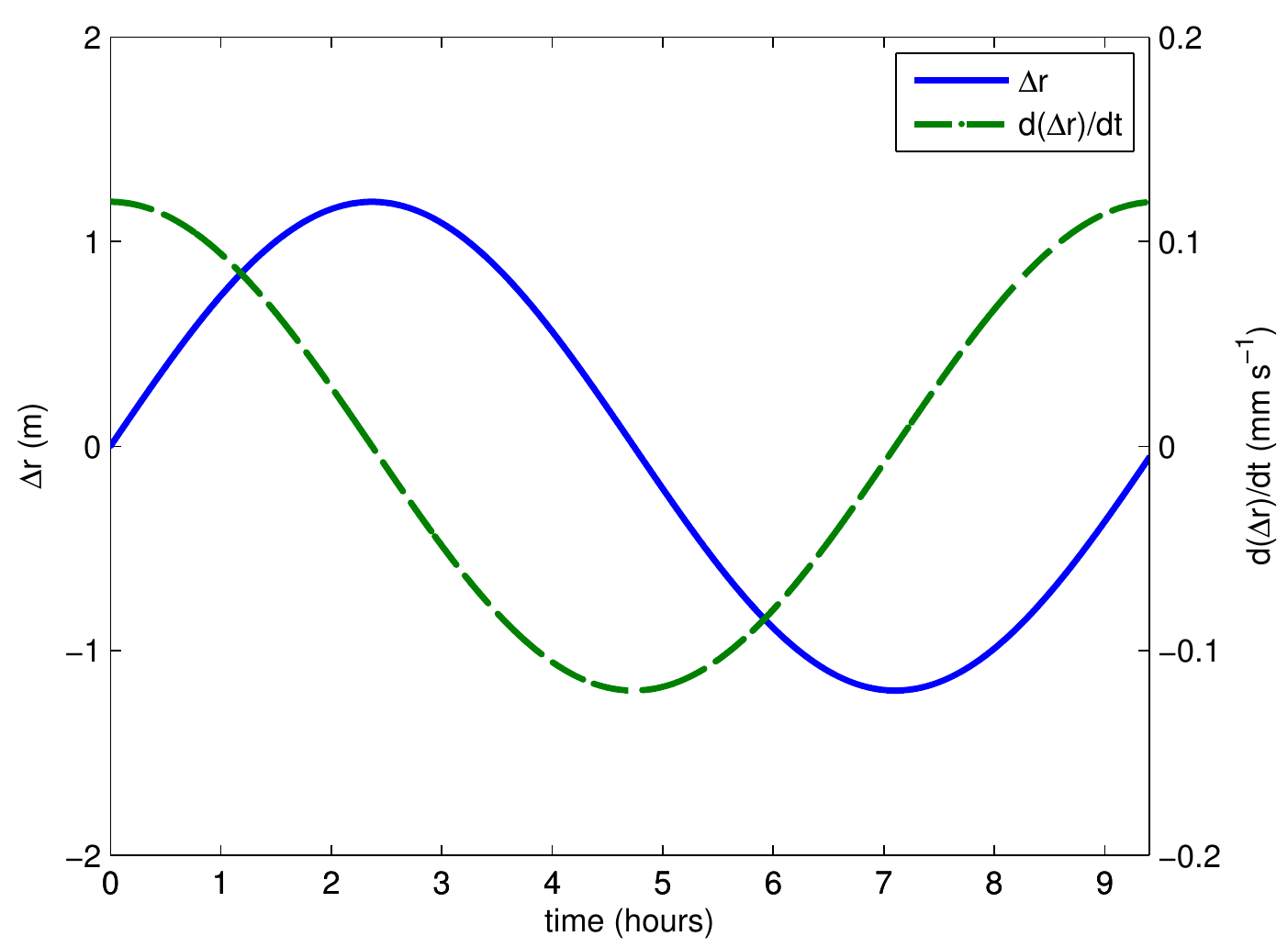}
 \caption{\csentence{Path length difference and velocity difference vs. time, due to motion of the satellite.} The solid blue line shows the free-space path length difference, and the dashed green line shows the rate of change of this difference. This is compensated for in post-processing, rather than in-flight.}
 \label{fig:path_length_compensation}
\end{figure}

Finally, the radial motion of the satellite causes constant variation in optical path length. To keep the path lengths of the interferometer equal this variation must be continuously compensated for. The time spent in the fibre loop by the part of the photon state in the upper arm, before it is emitted from the satellite, is given by:

\begin{equation}
\Delta t = \frac{\ell}{c n} \approx 10^{-4}\,\text{s}
\end{equation}
\noindent where $\ell = 60$\,km is the fibre length and $n=1.5$ is the refractive index. Multiplying this quantity by the radial velocity gives the change in path length between parts of the superposition due to the motion of the satellite; this change in path is plotted in Figure~\ref{fig:path_length_compensation}. However, this effect can be precisely calculated in advance and so either active compensation with optical components or passive compensation at the post-processing stage can be built into the experiment beforehand.

% Loss and link budget -------------------------------------------------------------------
\subsection{Loss}\label{sec:loss}

Given the repetition rate of the laser and strength of attenuation, we can calculate the expected rate of signal photons registering at the detectors. The analogous discussion on noise photons reaching the detectors is deferred to Section~\ref{sec:noise}. There are three major sources of signal attenuation: divergence of the beam from the transmitting telescope to the ground; loss  due to atmospheric irradiance and interference; and loss within the fibres.

% Atmospheric transmission ------------------------------------------------------
First, we examine the effects of atmospheric transmission. A wavelength of 1550nm is deemed suitable as it is standardised for telecoms use, maximises transmittivity in the fibre and is not readily absorbed by the atmosphere. The high frequency of the emitted photon allows us to ignore ionospheric effects on polarisation and is easily distinguishable from auroral activity over the ground stations \cite{Chamberlain1961}.

\begin{figure}
  %\hspace*{0.8cm}
  \includegraphics[width=0.8\linewidth, trim = 0.25cm 0.5cm 0.25cm 0cm]{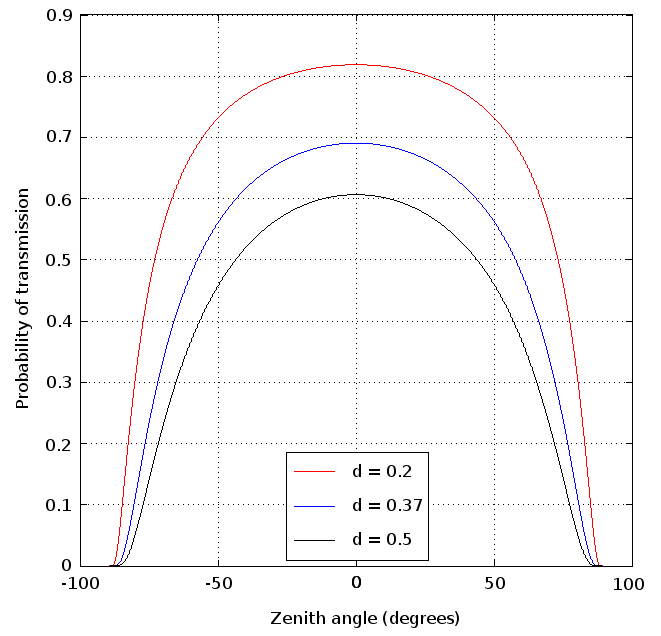}
\caption{\csentence{Probability of a photon to be transmitted through the atmosphere with respect to zenith angle, $\theta$, and various optical depths, d.} An optical depth, d, of 0.5 corresponds to high levels of pollution. Further calculations assume d=0.37, giving an atmospheric transmission probability of 0.69 at zenith.}
 \label{fig:PTrans_Zenith}
\end{figure}

Assuming homogeneity, using a variant of the Beer-Lambert Law we can estimate the probability of a single photon passing through the atmosphere as:

\begin{equation}
\text{P(transmission)} = \exp \left( \frac{-\tau_0}{\eta_0} \right)
\end{equation}
Where $\tau_0$ is the optical depth of the atmosphere, and $\eta_0$ is the angle of incidence of the beam. The optical depth varies over time and depends on myriad dynamic factors such as aerosol content, atmospheric mixing and Raman and Rayleigh scattering. Modelling these effects is essential, so one might consider optical depth readings from MODIS, MISR or future Sentinel satellites. A numerical weather prediction model might then be produced detailing daily optical depth over ground stations. An alternative might be to model turbulence transfer functions as in \cite{Sadot1995}. As shown in Figure \ref{fig:PTrans_Zenith}, we can allow for an optical depth of 0.5 before the satellite signal is severely attenuated. Combining these factors produces an estimate for the atmospheric transmission loss to be $\simeq -2.3\,\text{dB}$.

% Beam divergence ----------------------------------------------------------------
As for beam divergence, a downlink direction is chosen as accentuated divergence would then occur only in the last 12\,km of travel. This downlink beam divergence can be computed for a photon of wavelength $\lambda$ by:
\begin{equation}
\theta_{div} = \frac{\lambda}{\pi w_0} = 1.6\,\mu\text{rad}
\end{equation}
where $w_0 =$\,30\,cm corresponds to the radius of the transmitting telescope on the satellite. Consequently, the beam diameter on the ground at maximum and minimum altitude is given by $D_{ground} \simeq 2h\theta$, which implies that $58\,\text{m}\leq D_{ground}\leq 105\,\text{m}$, taking bounding values for the altitude as $18000\,\text{m} \leq h \leq 32000\,\text{m}$.

% Atmospheric turbulence ---------------------------------------------------------
Atmospheric turbulence must also be considered. As the atmospheric parameters used for post-processing are extracted from data from the reference laser, this turbulence must not radically change in the time between measurement of the reference laser and of the single photon source. Assuming that the gap between reference and single photon measurement is half the repetition rate of the laser, this delay comes to 5\,ns. Assuming a wind speed of 10\,m\,s$^{-1}$, an air parcel would move by about 50\,nm in some direction between the single photon measurement and reference laser measurement. This is considered negligible compared to other atmospheric effects.

% Link budget ---------------------------------------------------------------------
Atmospheric transmission loss, turbulence and beam divergence can be compiled into a ``link budget'' that calculates the full transmission loss from satellite to ground station. We can modify and simplify the Friis Transmission Equation to give the following link budget equation \cite{Shaw2013}:

\begin{equation}
\text{Loss}(dB) =10 \log \left[\left(\frac{\pi D_T D_R}{4 \lambda h}\right)^2 L_p L_t\right],
\end{equation}
where here $D_T$ is the transmitter diameter, $D_R$ is the receiver diameter, $\lambda$ is the wavelength of the single photon source, $h$ is the satellite altitude, $L_p$ is the pointing loss (taken to be 0.63 \cite{Timofeyev2008}), and $L_t$ the atmospheric transmission loss as calculated above.

Assuming an apogee of 32000\,km and a lowest altitude bin of 18000\,km, a ground receiver diameter of 3\,m, a satellite transmitter diameter of 0.6\,m, and superconducting detector efficiencies of about 90\%~\cite{Marsili2013}; we calculate the baseline signal gain from the link budget to be -30\,dB at perigee, and -35\,dB at apogee. 

Further factoring in losses from the optical fibre of -15.5 dB and fibre blackening from radiation of -1.7 dB, the link budget achieves a total attenuation of -52.2 dB in the worst case, towards the end of mission lifetime. Of course, to further reduce signal attenuation, one could increase the aperture diameters of the receiver and transmitter. However, this causes a loss in manoeuvrability and a significant cost increase for rapidly decreasing returns.

% Noise ----------------------------------------------------------------------------------
\subsection{Noise}\label{sec:noise}

A feasibility case must also ensure a signal-to-noise ratio (SNR) great enough to produce enough meaningful data for statistical analysis. It must be stressed that the signal received on Earth is fixed by the link budget above, however since we are trying to detect a single photon from a plethora of solar and planetary photon noise, optimising the SNR is crucial.

The effects of noise on space to ground quantum channels has previously been explored \cite{Er-long2005}. The noise power received by the ground telescope ($P_{\mathrm{b}}$) can be expressed as

\begin{equation}
\label{eq:Pbg}
P_{\mathrm{b}} = \frac{1}{4} H_{\mathrm{b}} \Omega_{\mathrm{fov}}  \pi D_R^{2}  \Delta \nu \Delta t_d\,,
\end{equation}
where $H_{\mathrm{b}}$ is the brightness of the sky in units of W\,m$^{-2}$\,sr$^{-1}$\,$\mathrm{\mu} $m$^{-1}$, $\Omega_{\mathrm{fov}}$ is the field of view, $\Delta \nu$ is the bandwidth and $\Delta t_d$ is the detection time. 

Typical sky brightness for quantum cryptography applications is discussed in \cite{Er-long2005}, using data from \cite{Leinert1998}. However, the data therein must be modified in light of the experiment proposed here. Firstly, the data presented in \cite{Er-long2005,Leinert1998} is for a frequency band just below 1550\,nm, which is subject to much less noise than at 1550\,nm itself. Conversely, the primary source of noise photons at 1550\,nm is hydroxyl airglow, the strength of which is strongly dependent on ambient temperature. The data from \cite{Leinert1998} assumes a receiving station in the tropics at Mauna Kea, whereas we propose an arctic station at Svalbard. Utilising instead polar sky brightness data from \cite{Phillips1999}, we take a sky brightness of 2$\times 10^{-5}\,$W$\,$m$^{-2}\,$sr$^{-1}\mathrm{\mu}$m$^{-1}$ at our operational frequency. The received signal band is filtered to a bandwidth of 15\,nm, corresponding to twice the bandwidth of the single photon pulses (twice the full width at half maximum of the Lorentzian pulse). The field of view of the receiving telescope is assumed to be 10$\,\mathrm{\mu}$rad, but the effective field of view is further reduced by a factor of 10 with a variable iris diaphragm \cite{Schmitt2007}. Noise power is further reduced with a 50$\,$ps time gate filter leading to an available detection time of 50$\,$ms per second. Reflections from the satellite or its black body radiation do not significantly contribute to the background noise \cite{Er-long2005}. Besides received background photons, total noise power also depends on the dark count rate of the single photon detectors. Superconducting detectors have negligible intrinsic dark count rate but as a worst case estimate the detector system dark count rate is assumed to be 1$\,$kHz (although a large fraction of these counts are neglected due to time gating) \cite{Marsili2013}. %Using these values the total noise power is  $P_{b}=1.9\,$aW, corresponding to a noise photon rate of $N_{\mathrm{b}}\approx$15$\,$s$^{-1}$. 
The rate of detected signal photons is calculated as 590\,Hz (using the repetition rate of the laser, attenuation and loss). Combining this figure with the expected noise from ambient photons and system dark count gives an SNR $\approx 9.0 = $9.6$\,$dB.
%\begin{align}
%\nonumber SNR &= \frac{\text{Signal photons}}{\text{Noise photons}}\\
%\nonumber &= 10\log\left(\frac{4092.6}{1.25\times10^{-5}}\right)\\
%&= 85.15 dB.
%\end{align}

%---------------------------------------------------------------------------

\section{Hypothesis testing}\label{sec:hypothesis_testing}

Once an experimental profile has been outlined, we can perform a statistical analysis to extract the confidence with which we can confirm, or refute, the hypothesis in Eq.~\ref{eq:MZ} (and likewise, the theoretical analysis presented in \cite{Zych2012}).

We assume data taken in $k$ altitude bins, uniformly spaced in $h$ from $h_{min} =$\,18000\,km to $h_{max} =$\,32000\,km, with $n$ counts per bin. Then, we take normally distributed experimental parameters with mean as given in the specification in Section~\ref{sec:setup}, and variance as fixed by the discussion on stability in Section~\ref{sec:stability}. We assume that a fraction of the photons are lost according to the link budget in Section~\ref{sec:loss}, and that there is an ambient background of noise photons as per Section~\ref{sec:noise}. We then perform Monte Carlo simulations for a range of number of altitude bins, $k$, and counts per altitude bin, $n$, to generate artificial data sets given this initial prescription.

To extract a statistical significance from these simulations, we perform a Kolmogorov-Smirnov statistical significance test between the simulated data set and the mathematical formulation in Eq.~\ref{eq:MZ} and \cite{Zych2012}. The test is devised to extract the goodness of fit for a data set to an arbitrary functional hypothesis. Using this test, we can predict the number of single photons that must be sent in order to test the hypothesis to a specified confidence.

\begin{figure}[t!]
  \centering
 \includegraphics[width=0.85\linewidth,trim = 1.0cm 0.7cm 1.2cm 0.7cm]{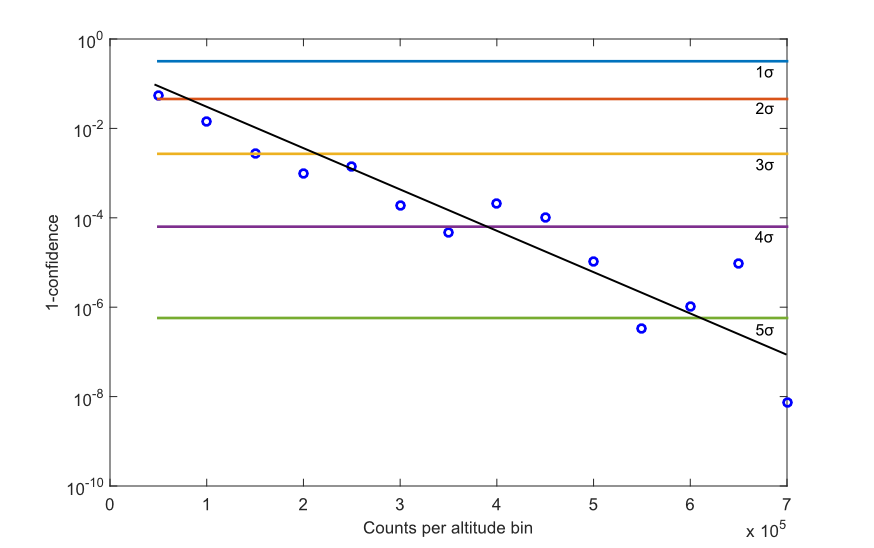}
 \caption{\csentence{Monte Carlo estimation of the number of counts per altitude bin required in order to confirm (or refute) Eq.~\ref{eq:MZ} to a specified confidence.} Horizontal lines indicate increasing "sigmas" of confidence, descending from $1\sigma$ to $5\sigma$.}
 \label{fig:hyp_test}
\end{figure}

The consequential data from the simulation and statistical test is shown in Figure~\ref{fig:hyp_test}, for $k=500$ altitude bins. Here, we can see that the number of received, true counts per altitude bin must exceed $\approx 6.5 \times 10^5$ counts in order to test the hypothesis to a $\gtrapprox 5\sigma$ confidence. Given the figure for total loss of $-52.2$\,dB, this amounts to a necessary emission total of $\approx 10^{11}$ single photons per altitude bin. Given the repetition rate of the laser and the degree of attenuation, we expect raw emission rates of 100\,MHz and thus reach the required emission total in each altitude bin within $\sim 1000\,\text{ seconds } = 16.7\,\text{ minutes}$ of continuous operation.

%--------------------------------------------------------------------------

\section{Mission Design}\label{sec:mission_design}
%Virginia is in charge

	In order to reach the desired confidence defined in Section~\ref{sec:hypothesis_testing}, data must be taken over a sufficiently large difference in gravitational potential. An orbit with a perigee of 700\,km and an apogee of 32000\,km gives access a relativistic time delay of 150\,fs, which is both large enough to be resolved by the detectors and still gives a reasonable count rate at the apogee. During the measurement procedure, every other light source introduces noise. The ambient sunlight is indeed strong enough to wash out the signal from the satellite. Therefore, measurements must be performed when the ground station is not illuminated by the Sun. Additionally, in order to have the maximum number of measurements per orbit and to minimise noise photons from airglow, the ground station needs to be placed as close as possible to the North Pole. 
	
	These considerations lead to the selection of the ground station located in Svalbard, Norway.
	By considering the maximum optical path that gives a valid measurement and the movement capabilities of the ground telescope, a connection cone of 45\textdegree \,around the zenith can be defined. This indeed allows satellite access times for up to 7 hours per orbit. This ground station is ideal for the mission, though it introduces a constraint on orbital inclination. In fact, only a polar orbit can provide the maximum visibility time from the ground.
	
\begin{figure}
		\includegraphics[width=0.8\linewidth, trim = 2cm 2cm 0cm 0cm]{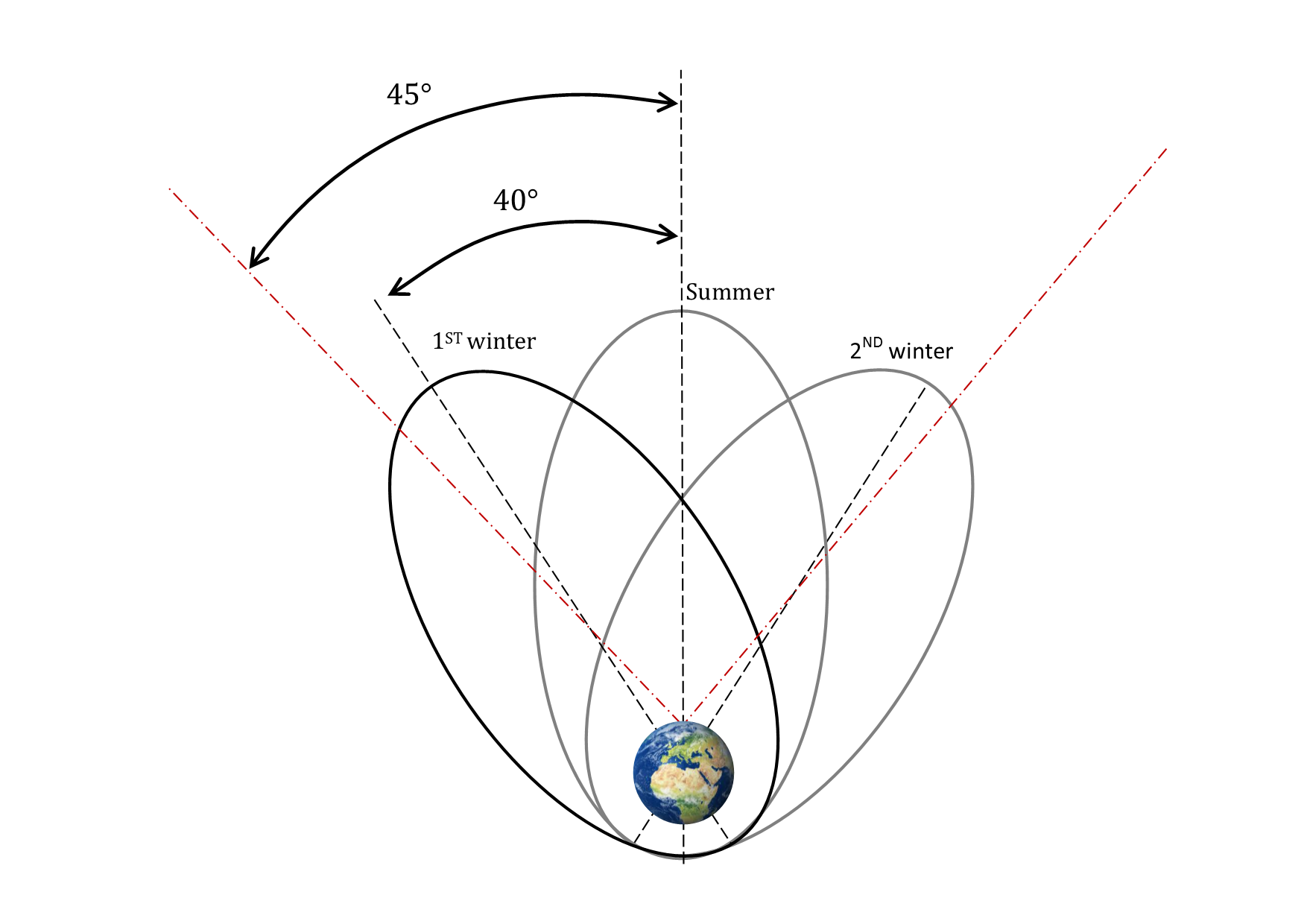}
		\caption{\csentence{Schematic of mission phases and orbit drift.} Note the division into first and second operational phases.}
		\label{orbit}
	\end{figure}

	 The range of orbital heights chosen to send single photons is from 18000\,km to 32000\,km (to minimise special relativistic effects, as discussed in Section~\ref{sec:time_dilation}). Given the number of altitude bins and length of measurement windows, we estimate that a mission lifetime of 1.5 years will completely satisfy the previously stated requirements. The mission profile is thus composed of three different phases. The first, starting immediately after launch, is a 6-month commissioning period used to calibrate the orbit and the measurement system with laser ranging and radio communication. During the satellite motion, due to the perturbation from the oblateness of the Earth and the high eccentricity of the orbit, the orbital apse line rotates clockwise with a rate of $77.1\degree$/year. With an initial argument of perigee of $\omega = 350\degree$ and a launch during Svalbard's spring season, after the commissioning phase $\omega \simeq 310\degree$, meaning that the apogee is contained in the ground station visibility cone. This situation is represented by the thicker line in Figure~\ref{orbit}.  
The main operational phase thus begins in winter and lasts for approximately 6 months. It is then followed by a second mission phase in Svalbard's summer, during which the first results are examined and the orbit is further calibrated. Finally, the last 6-month phase provides additional measurements. At the end of the 1.5-year lifetime, a 55\,ms$^{-1}$ thruster burn lowers the perigee to approximately 200\,km. This manoeuvre eventually leads to a further lowering of the apogee and then to a controlled de-orbit into the atmosphere.

The total satellite mass is approximated at 400\,kg. The small spacecraft size allows using the VEGA launcher in order to bring the satellite into a parking orbit with a perigee of 700\,km and an apogee of 20000\,km. The final orbit is then reached using an on board bi-propellant propulsion system through a perigee burn with a $\Delta V$ of 323\,ms$^{-1}$ .

\section{Risk Analysis}\label{sec:risk}

Here we discuss risk control and mitigation during the mission.

\subsection{Radiation effects}
The spacecraft's immediate radiation environment, composed of fluxes from the solar wind and galactic cosmic rays, has a non-negligible effect on the performance of the components. During its orbit around Earth, the satellite passes twice through the Van Allen belts. Therefore, suitable shielding is necessary to ensure the correct operation of the payload. Particular precaution has to be taken to shield the spools of optical fibre, due to the effect of Radiation-Induced Attenuation (RIA). Single mode fibres with an RIA of 0.5\,dB/km per 9000\,rad have been demonstrated in the literature \cite{Uffelen2002}, a figure which can be used for a worst-case analysis. Assuming a 20\,mm thick spherical layer of aluminium shielding, the fibre receives an estimated Total Ionizing Dose (TID) between 500 and 2000\,rad/year, which yields an RIA between -1.7 and -6.7\,dB/year. 

\begin{figure}
	\centering
	\includegraphics[width=0.9\linewidth,trim=0.7cm 0.5cm 0.7cm 0.8cm]{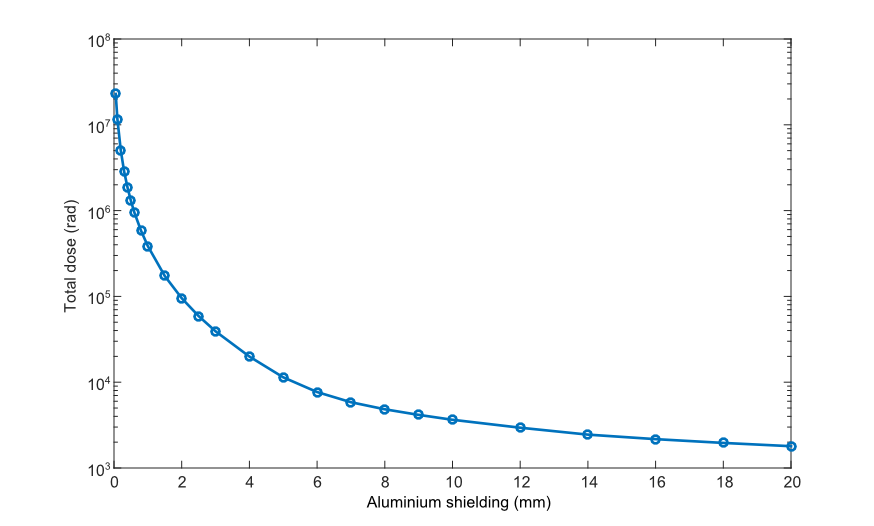}
	\caption{\csentence{Total Ionizing Dose as a function of the Aluminium shielding.} Choice of shielding thickness is a compromise between continued performance of the optical fibres and payload weight.}
\end{figure}

An aluminium shielding of 2\,mm is provided for the lasers and the optical bench. This is also ample shielding to guarantee low radiation doses for the remaining optics. 

The blackening of fibre cables due to radiation is thus one of the main operational risks. Although this process is slowed by adequate shielding, occurrence at a higher than expected rate would pose a major obstacle to collecting adequate data. Active avoidance of the Van Allen belts by modifying the orbit could be pursued, if needed.

\subsection{Thermal Control System}
During the mission, the satellite undergoes cyclic eclipses; the resulting temperature fluctuations have to be analysed in order to design the thermal control system. The most critical requirement is the temperature stability of the optical bench. Insulation ensures that the fibre temperature remains within 0.5\,K of the equilibrium value. Specifically, the satellite is equipped with a Multi Layer Insulation (MLI) aluminium and Kapton coating and heat pipes to convey and redistribute heat. With the appropriate sizing for radiators and heat pipes it is possible to achieve the required thermal stability, although this represents another primary risk for the mission.

\subsection{Attitude and Orbit Control System}

Precise determination of the satellite's orbit and fine pointing are also fundamental. 

Firstly, we assumed in Sec~\ref{sec:loss} that the satellite's pointing accuracy is $0.5\,\mu$rad, with $0.1$\,mrad/s maximum slew rate. This strict requirement ensures efficient transmission of the required number of photons to the ground, but is a relevant risk for the mission. Indeed, if the satellite failed on maintaining the required accuracy, a prohibitively large fraction of the signal photons would be lost. The satellite is thus equipped with a tracking telescope to have a first estimate of the attitude; then, two star trackers refine the determination. A system of reaction wheels and vibration dampers allows for a fine control of the satellite orientation. 

As mentioned above, orbital determination is of key importance for the mission. Indeed, in order to account for the transmission error due to the radial motion of the satellite, the radial velocity must be known to within $\sim 1$\,mm/s precision. This is feasible with current radio tracking systems working at either S- or X-band. In fact, the commonly used Ultra-Stable Oscillators often have a Van Allen stability better than 10$^{-13}$ over an integration time from 10\,s to 1000\,s, which is completely adequate to fulfil the requirement.

\section{Conclusion}

We have demonstrated herein a proposal capable of probing the interplay between quantum mechanics and general relativity using single photon interferometry - in particular, exploring the dichotomous nature of time in the two theories. Moreover, the apparatus is well within the reach of current quantum optics technology, and there is already historical precedent for space-qualification and launch of similar payloads. Conversely, successful operation of this experiment would provide strong precedent itself for future projects involving commercial quantum communications satellites.

%%%%%%%%%%%%%%%%%%%%%%%%%%%%%%%%%%%%%%%%%%%%%%
%%                                          %%
%% Backmatter begins here                   %%
%%                                          %%
%%%%%%%%%%%%%%%%%%%%%%%%%%%%%%%%%%%%%%%%%%%%%%

\begin{backmatter}

\section*{Competing interests}
  The authors declare that they have no competing interests.

\section*{Author's contributions}
    SP was primary editor of the paper. The introduction and scientific background were written by SP, VF, PK and RA. The implementation and setup was primarily devised by SC and NB. Derivations and implementations regarding transmission loss, noise and atmospheric effects were written by NG, ML and J-P K. The hypothesis testing and Monte Carlo simulations were written by SP. The mission design and engineering sections were written by VN, SdC and TN. All other authors were contributory members during the Alpbach and Post-Alpbach conferences.

\section*{Acknowledgements}
  SP was supported by the Bristol Quantum Engineering Centre for Doctoral Training, EPSRC grant EP/L015730/1. PK was supported by the UK EPSRC through the Quantum Technology Hub: Networked Quantum Information Technology (grant reference EP/M013243/1). NG was supported by the Groupement Luxembourgeois de l'Aéronautique et de l'Espace (grant GLAE-Alpbach). This work was carried out primarily at the Alpbach and Post-Alpbach conferences. The Summer School Alpbach: Sixty  highly qualified European science and engineering students converge annually for stimulating ten days of work in the Austrian Alps. Four teams are formed who each design a space mission which are then judged by a jury of experts. Students learn how to approach the design of a satellite mission and explore new and startling ideas supported by experts. The Summer School Alpbach enjoys 40 years of tradition in providing in-depth teaching on different topics of space science and technology, featuring lectures and concentrated working sessions on mission studies in self-organised working groups. The Summer School Alpbach is organised by FFG and co-sponsored by ESA, ISSI and the national space authorities of ESA member and cooperating states. We would like to specifically acknowledge the tutelage of G\"unter Kargl, Manuela Werner, Siddharth Koduru Joshi, Esteban Castro-Ruiz and Rupert Ursin, whose input was instrumental.
%%  After submission of the .TEX file,                     %%
%%  you will be prompted to submit your .BBL file.         %%

% if your bibliography is in bibtex format, use those commands:
\bibliographystyle{bmc-mathphys} % Style BST file (bmc-mathphys, vancouver, spbasic).
\bibliography{JANOS.bib}      % Bibliography file (usually '*.bib' )

\end{backmatter}
\newpage
\appendix

\section{Derivation of the relativistic time delay}\label{appendix:deltatau}

Here we show how to derive Eq. (\ref{eq:tdelay}), generalising the calculation described in \cite{Zych2012}.

First of all, we identify the reference frame of a distant observer with the ECI (Earth-Centered Inertial) frame. In order to take into account the velocity of the satellite and of the ground station rotating with the Earth's surface, Eq. (7) in \cite{Zych2012} must be modified. Let $\tau_r$ and $t$ be the proper time recorded by a clock at the radial distance $r$ and the coordinate time recorded by a distant observer, respectively. Then we have:
\begin{equation}
d\tau_r^2=\left(1+\frac{2V(r)}{c^2}-\frac{v_r^2}{c^2}\right)dt^2
\label{eq:metric}
\end{equation}
where $V(r)=-\frac{GM}{r}$ is the Earth's gravitational potential and $v_r$ is the speed of the clock at the radial distance $r$, as measured in the ECI frame.

If the fibre refractive index is $n$ and $c$ is the speed of light in vacuum, the proper time spent by a photon in a fibre loop of proper length $l$ is $nl/c$. Hence, using Eq. (\ref{eq:metric}), we can find the time spent by a photon in the two fibres, as measured by a distant observer:
\begin{equation}
t_r=\frac{nl_r}{c\sqrt{1+\frac{2V(r)}{c^2}-\frac{v_r^2}{c^2}}}
\label{eq:tr}
\end{equation}
where $l_r$ is the proper length of the fibre at the radial distance $r$. Henceforth, the quantities related to the satellite and to the ground station will be indicated by the subscripts $s$ and $g$, respectively. We assume $l_s=l$ and $l_g=l+dl$.

The difference in photon arrival times (referred to as the ``time delay'') measured by a distant observer is $t_s-t_g$. Hence, using Eq. (\ref{eq:metric}) and (\ref{eq:tr}), the time delay measured by the local observer at the ground station is:
\begin{equation}
\begin{aligned}
\Delta\tau &=\sqrt{1+\frac{2V_g}{c^2}-\frac{v_g^2}{c^2}}(t_s-t_g)\\
&=\frac{nl}{c}\left(\frac{\sqrt{1+\frac{V_g}{c^2}-\frac{v_g^2}{c^2}}}
{\sqrt{1+\frac{V_s}{c^2}-\frac{v_s^2}{c^2}}}-1\right)-\frac{ndl}{c} .
\label{eq:delaystep1}
\end{aligned}
\end{equation}
Taking the first order approximation to the potential and the second order approximation to the velocity, Eq. (\ref{eq:delaystep1}) becomes:
\begin{equation}
\Delta\tau=\frac{nl}{c^3}\left(V_g-\frac{v_g^2}{2}-V_s+\frac{v_s^2}{2}\right)-\frac{ndl}{c} .
\label{eq:delaystep2}
\end{equation}
Assuming that the ground station is at rest on the geoid, we can use the conventional geoidal potential $W_0$ to write the time delay of Eq. (\ref{eq:delaystep2}) as:
\begin{equation}
\Delta\tau=\frac{nl}{c^3}\left(-W_0-V_s+\frac{v_s^2}{2}\right)
-\frac{ndl}{c} .
\label{eq:delaystep3}
\end{equation}
Finally, for an elliptical orbit with semi-major axis $a$, indicating with $h$ the altitude of the satellite on the Earth's surface, the time delay of Eq. (\ref{eq:delaystep3}) becomes:
\begin{equation}
\Delta \tau=\frac{nl}{c^3}\left[-W_0+GM\left(\frac{2}{R_{\oplus}+h}-
\frac{1}{2a}\right)\right]-\frac{ndl}{c}
\label{eq:delayfinal}
\end{equation}
which is exactly Eq. (\ref{eq:tdelay}) in the text.

\section{Derivation of the expected interferometric signal}\label{appendix:signal}

Here we show how to obtain Eq. (\ref{eq:MZ}) in the text. 

The commutation relation for continuous mode photon states can be written \cite{Loudon2003}:
\begin{equation}
[a_{\nu'},a_\nu^\dagger] = \delta(\nu - \nu').
\label{eq:commutation_relation_1}
\end{equation}

Say we have a Mach-Zehnder interferometer with equal arm lengths, but with one arm subject to time-dilation; then the state at detector $D\pm$ with a single input photon is:
\begin{equation}
\nonumber | 1 \rangle _{\nu \pm} \propto \int d \nu f(\nu) \left( e ^{i \frac{\nu}{c}(x_r - c \tau_r)} \pm
e ^{i \frac{\nu}{c}(x_r - c (\tau_r + \Delta \tau))} \right)
  a^\dagger_\nu |0\rangle.
\end{equation} 
Where $x_{r}$ is the local Cartesian coordinate (perpendicular to the radial coordinate $r$), , $\tau_r$ is the local time coordinate, $f(\nu)$ is the spectrum of the light source, $\nu$ is angular frequency, $c$ is the speed of light, and $\Delta\tau$ is as given in Eq. (\ref{eq:tdelay}).

The probability of a photon arriving at this detector is then
\begin{equation}
\begin{aligned}
P _\pm &=  _{v' \pm}\langle 1 | 1 \rangle_{\nu \pm} \\
&\propto \langle 0 | a_{\nu'} 
\int d \nu \int d \nu'
f(\nu) f(\nu')\\
&\times \left( e^{i \frac{\nu}{c}(x_r - c \tau_r)} +
e^{i \frac{\nu}{c}(x_r - c ( \tau_r + \Delta \tau)} \right) \\
& \times \left(e^{-i \frac{\nu'}{c}(x_r - c \tau_r)} +
e ^{-i \frac{\nu'}{c}(x_r - c ( \tau_r + \Delta \tau)} \right) a^\dagger_\nu |0\rangle \\
\label{eq:prob1}
\end{aligned}
\end{equation}
From commutation relation Eq. (\ref{eq:commutation_relation_1}), it is shown that
\begin{equation}
_{\nu'} \langle 1 | 1 \rangle _{\nu} = \delta(\nu - \nu')
\end{equation}
Thus
\begin{equation}
\begin{aligned}
P _\pm &\propto \int d \nu \int d \nu'
\left( e ^{i \frac{\nu}{c}(x_r - c \tau_r)} +
e ^{i \frac{\nu}{c}(x_r - c ( \tau_r + \Delta \tau)} \right) \\
& \times f(\nu) f(\nu') \left( e ^{-i \frac{\nu'}{c}(x_r - c \tau_r)} +
e ^{-i \frac{\nu'}{c}(x_r - c ( \tau_r + \Delta \tau)} \right) \\
&\times \delta(\nu - \nu') \\
&= \int d\nu |f(\nu) \left(e^{i \frac{\nu}{c}(x_r - c \tau_r)} + e ^{i \frac{\nu}{c}(x_r - c ( \tau_r + \Delta \tau)} \right)|^2 \\
&= \frac{1}{2} \left( 1 + \int d\nu |f(\nu)|^2 \textrm{cos}(\nu \Delta \tau) \right).
\label{eq:prob2}
\end{aligned}
\end{equation}

\end{document}